\documentclass[journal]{IEEEtran}
\usepackage{graphicx}      % include this line if your document contains figures
\usepackage{subfig}
\usepackage[square,sort,comma,numbers]{natbib}
\usepackage{amsmath}
\usepackage{amsfonts}
\usepackage{tablefootnote}
\usepackage{url}
\usepackage[table]{xcolor}
\usepackage{color}
\usepackage{bm}

\usepackage{nth}
\usepackage{tikz}
\usepackage{verbatim}
\usepackage{tkz-euclide}

\definecolor{ao(english)}{rgb}{0.0, 0.5, 0.0}

\usepackage[utf8]{inputenc}

\begin{document}
\title{Coupled and decoupled impedance models compared in power electronics systems\\}

\author{{Atle~Rygg,~Marta~Molinas,~Chen~Zhang~and~Xu~Cai}%

\thanks{A. Rygg and M. Molinas are with the Dept. of Engineering Cybernetics, Norwegian University of Science and Technology, O. S. Bragstads plass 2D, 7032 Trondheim, Norway, e-mail: atle.rygg@itk.ntnu.no, tel: +47 95977295}%
\thanks{C. Zhang and X. Cai are with the State Energy Smart Grid R\&D Center, Shanghai Jiao Tong University, Shanghai, China, e-mail: zhangchencumt@163.com}}

\maketitle

\begin{abstract}             
This paper provides a comparative analysis of impedance models for power electronic converters and systems for the purpose of stability investigations. Such models can be divided into either decoupled models or matrix models. A decoupled impedance model is highly appealing since the Single-Input-Single-Output (SISO) structure makes the analysis and result interpretation very simple. On the other hand, matrix impedance models are more accurate, and in some cases necessary. Previous works have applied various approximations to obtain decoupled models, and both the \textit{dq}- and sequence domains have been used. This paper introduces the terms \textit{decoupled} and \textit{semi-decoupled} impedance models in order to have a clear classification of the available approximations.

The accuracy of 4 decoupled impedance models are discussed based on the concept of Mirror Frequency Coupling (MFC). By definition the decoupled models based on sequence domain impedances will be exact for systems without MFC. In the general case, they are expected to be more accurate than the decoupled \textit{dq}-impedance models. The paper defines a norm $\epsilon$ to measure the degree of coupling in the impedance matrices. This norm equals the error in the eigenvalue loci between the matrix and semi-decoupled models. This can also be viewed as the error in the semi-decoupled Nyquist plot.

An example case study consisting of a grid-connected VSC with current controller and PLL is used to compare the different methods. It is found that decoupled and semi-decoupled models in the \textit{dq}-domain are only applicable in grids with very low X/R-ratio. Furthermore, it is concluded that the decoupled model in the sequence domain gives close to equal results as the semi-decoupled model.

\end{abstract}


% \fi
%\IEEEpeerreviewmaketitle

\section{Introduction}
Systems that are Time Invariant (TI) in the synchronous rotating reference frame (\textit{dq}-domain) can be modeled accurately by an impedance matrix. This was first performed in \cite{Belkhayat1997}. Another track of research has applied harmonic linearization by symmetric components in the phase domain \cite{Sun2011}. The latter representation does not use matrix models, and is therefore viewed as \textit{decoupled} by the definitions in this paper. It was recently shown that the \textit{dq}-domain impedance matrix has an equivalent matrix in the sequence domain where positive and negative sequence are shifted by two times the fundamental frequency. The term \textit{modified sequence domain} is defined in order to derive this equivalence \cite{Rygg2016}. 

This paper investigates the possible methods for approximating a decoupled impedance model. The terms \textit{decoupled} and \textit{semi-decoupled} models are defined to clear an important ambiguity. A decoupled model neglects the off-diagonal elements initially, while the semi-decoupled models neglects the off-diagonal elements after the full 2x2 matrices are obtained. The upside of decoupled models is their simplicity when extracting models from simulations or experiments.  

Section \ref{sec:models} contributes with the definitions and classifications of impedance models, as well as an overview of previous works. In addition, the expressions for minor-loop gains and eigenvalue loci are given. Section \ref{sec:MFD} discusses the role of Mirror Frequency Coupling (MFC) in decoupled impedance models, while section \ref{sec:inj} presents the various perturbation injection signals applied in this work. Finally, simulation results are used to illustrate the different impedance models which are discussed in light of the norm $\epsilon$ in section \ref{sec:sim}.

\section{Overview of impedance models}\label{sec:models}
This paper assumes systems that are time invariant in the \textit{dq}-domain. This is a simplification that neglects certain effects such as power electronic switching. The assumption is widely applied in stability analysis of power electronic systems, and will make interpretation of results much easier. The following three terms are defined in both impedance domains:

\begin{itemize}
\item \textit{Decoupled models} initially assumes two independent Single-Input-Single-Output (SISO)-models, i.e. neglects all coupling
\item \textit{Semi-decoupled models} captures all coupling by a Multiple-Input-Multiple-Output (MIMO)-model, but neglects the resulting coupling at the final stage
\item \textit{Exact models} represents the system by 2x2 matrices, and performs stability analysis by MIMO-methods
\end{itemize}

\subsection{Matrix (exact) impedance models}

Under the assumption of time invariance in the \textit{dq}-domain, the following model can be used to accurately describe the small-signal dynamics of a power electronic system \cite{Belkhayat1997}:

\begin{equation}
\mathbf{Z}_{dq}=
\begin{bmatrix}
Z_{dd} & Z_{dq} \\ Z_{qd} & Z_{qq}
\end{bmatrix}
\qquad
\mathbf{Y}_{dq}=\mathbf{Z}_{dq}^{-1}=
\begin{bmatrix}
Y_{dd} & Y_{dq} \\ Y_{qd} & Y_{qq}
\end{bmatrix}
\label{eq:Zdq_full}
\end{equation}

In the present paper the word \textit{exact} will be used to describe these models, since they they do not introduce any error under the time invariance assumption. 

In \cite{Rygg2016} the modified sequence domain was defined as an extension to the well established harmonic linearization in the sequence domain \cite{Sun2011}. By defining positive and negative sequence at the so-called mirror frequencies, the following impedance/admittance model was defined: 

\begin{align}
\mathbf{Z}_{pn}&=
\begin{bmatrix}
Z_{pp} & Z_{pn} \\ Z_{np} & Z_{nn}
\end{bmatrix} 
\qquad
\mathbf{Y}_{pn}=
\begin{bmatrix}
Y_{pp} & Y_{pn} \\ Y_{np} & Y_{nn}
\end{bmatrix} 
\nonumber \\
\omega_p&=\omega_{dq}+\omega_1 \nonumber \\
\omega_n&=\omega_{dq}-\omega_1
\label{eq:Zpn_full}
\end{align}

where $\omega_p$ and $\omega_n$ are the positive and negative sequence frequencies, respectively. Compared with the corresponding \textit{dq}-domain frequency $\omega_{dq}$, they are shifted by the fundamental frequency $\omega_1$ in opposite directions.

It was shown in \cite{Rygg2016} that $\mathbf{Z}_{dq}$ relates to $\mathbf{Z}_{pn}$ by

\begin{align}
\mathbf{Z}_{pn}&=A_Z \mathbf{Z}_{dq} A_Z^{-1} \nonumber \\
\mathbf{Y}_{pn}&=A_Z \mathbf{Y}_{dq} A_Z^{-1} \nonumber \\
A_{Z}&=\frac{1}{\sqrt{2}} 
\begin{bmatrix}
1 & j \\1 & -j
\end{bmatrix} \label{eq:imp_transf}
\end{align}

where the transformation matrix $A_Z$ is unitary since its inverse is equal to its complex conjugate transpose. It is known from linear algebra that eigenvalues are invariant when multiplied with unitary matrices as in (\ref{eq:imp_transf}). Consequently, the Nyquist plots in the \textit{dq}-domain and the modified sequence domain will be identical. 

When obtaining the impedance matrices in (\ref{eq:Zdq_full}) and (\ref{eq:Zpn_full}) from simulation or measurement, it is necessary to combine two linear independent injections as explained in \cite{Francis2011}.

\subsection{Decoupled impedance models}
Decoupled impedance models are obtained by completely neglecting coupling in the impedance matrices (\ref{eq:Zdq_full})-(\ref{eq:Zpn_full}). In the sequence domain this method is called \textit{harmonic linearization} since impedance is obtained by superimposing harmonic components onto the fundamental alternating waveforms \cite{Sun2011}. The following expressions can be used to find the decoupled impedance/admittance in the sequence domain:
\begin{align}
&\mathbf{Z}_{pn,\textrm{dec}}=
\begin{bmatrix}
Z_{p} &  0 \\ 0 & Z_{n}
\end{bmatrix} \qquad
\mathbf{Y}_{pn,\textrm{dec}}=
\begin{bmatrix}
Y_{p} &  0 \\ 0 & Y_{n}
\end{bmatrix}
\nonumber \\
Z_p&=\frac{1}{Y_p}=\frac{V_p}{I_p} \qquad
Z_n=\frac{1}{Y_n}=\frac{V_n}{I_n} 
\label{eq:Zpn_dec}
\end{align}

By using this impedance model, it is no longer needed to combine two linear independent injections in order to establish a full matrix. It was shown in \cite{Rygg2016} that the decoupled impedance equivalents depend on the injection method (e.g. shunt vs. series), and are consequently not uniquely defined. The sufficient condition for avoiding this ambiguity is to require Mirror Frequency Decoupled (MFD) systems \cite{Rygg2016}. The relation between $\mathbf{Z}_{pn,dec}$ and $\mathbf{Z}_{pn}$ was derived in the appendix of the same reference.

Decoupled impedance models in the \textit{dq}-domain have not been applied in previous research, but are defined in this paper to complete the systematic overview of models. The model is defined in a similar way as (\ref{eq:Zpn_dec}):

\begin{align}
&\mathbf{Z}_{dq,\textrm{dec}}=
\begin{bmatrix}
Z_{d} &  0 \\ 0 & Z_{q}
\end{bmatrix} \qquad
\mathbf{Y}_{dq,\textrm{dec}}=
\begin{bmatrix}
Y_{d} &  0 \\ 0 & Y_{q}
\end{bmatrix}
\nonumber \\
Z_d&=\frac{1}{Y_d}=\frac{V_d}{I_d} \qquad
Z_q=\frac{1}{Y_q}=\frac{V_q}{I_q} 
\label{eq:Zdq_dec}
\end{align}

\subsection{Minor-loop gains}\label{sec:minorloop}
The minor-loop gain $\mathbf{L}$ is needed to apply the Nyquist Criterion (NC) or the Generalized Nyquist Criterion (GNC) \cite{Desoer1980}. For matrix impedance models, the minor-loop gain is defined as:

\begin{equation}
\mathbf{L}=\mathbf{Z}_S \mathbf{Z}_L^{-1} = \mathbf{Z}_S \mathbf{Y}_L 
\label{eq:L}
\end{equation}

where $\mathbf{Z}_S$ is the source impedance and $\mathbf{Y}_L$ is the load admittance. Of note, recent works have proposed to apply the \textit{inverse} Generalized Nyquist Criterion in certain cases where GNC is hard to interpret due to open-loop unstable poles \cite{Wen2016b}. These special cases are not a topic of the present paper.

The minor-loop gains can be defined based on the impedance models as follows:

\begin{align}\label{eq:Ldq_exact}
&\mathbf{L}_{dq,\textrm{exact}} = \begin{bmatrix}
Z_{dd}^S Y_{dd}^L + Z_{dq}^S Y_{qd}^L & Z_{dd}^S Y_{dq}^L + Z_{dq}^S Y_{qq}^L \\ 
Z_{qd}^S Y_{dd}^L + Z_{qq}^S Y_{qd}^L & Z_{qd}^S Y_{dq}^L + Z_{qq}^S Y_{qq}^L 
\end{bmatrix}  \\\label{eq:Lpn_exact}
&\mathbf{L}_{pn,\textrm{exact}} = \begin{bmatrix}
Z_{pp}^S Y_{pp}^L + Z_{pn}^S Y_{np}^L & Z_{pp}^S Y_{pn}^L + Z_{pn}^S Y_{nn}^L \\ 
Z_{np}^S Y_{pp}^L + Z_{nn}^S Y_{np}^L & Z_{np}^S Y_{pn}^L + Z_{nn}^S Y_{nn}^L 
\end{bmatrix} \\ \label{eq:Ldq_dec}
&\mathbf{L}_{dq,\textrm{dec}} = \begin{bmatrix}
Z_{d}^S Y_{d}^L & 0 \\ 
0 & Z_{q}^S Y_{q}^L 
\end{bmatrix} \\
&\mathbf{L}_{pn,\textrm{dec}} = \begin{bmatrix}
Z_{p}^S Y_{p}^L & 0 \\ 
0 & Z_{n}^S Y_{n}^L 
\end{bmatrix} \label{eq:Lpn_dec}
\end{align}

where the subscript \textit{exact} is used to underline that the matrix models will capture all system dynamics without error under the time invariance assumption made in this paper.

\subsection{Semi-decoupled models}
The previous subsection presented a method to approximate decoupled impedance models. Another method for the same purpose has been applied in previous works for the \textit{dq}-domain \cite{Burgos2010} \cite{Wen2015b}. This method is based on first obtaining the complete minor-loop gain (\ref{eq:Ldq_exact})-(\ref{eq:Lpn_exact}), and then neglecting the resulting off-diagonal elements:

\begin{align}
\label{eq:Ldq_semidec}
&\mathbf{L}_{dq,\textrm{semidec}} =
\begin{bmatrix}
L_{dd} & 0 \\0 & L_{qq}
\end{bmatrix}\nonumber \\
&=
\begin{bmatrix}
Z_{dd}^S Y_{dd}^L + Z_{dq}^S Y_{qd}^L  & 0 \\ 
0 & Z_{qd}^S Y_{dq}^L +  Z_{qq}^S Y_{qq}^L 
\end{bmatrix} \\
&\mathbf{L}_{pn,\textrm{semidec}} = 
\begin{bmatrix}
L_{pp} & 0 \\0 & L_{nn}
\end{bmatrix}\nonumber \\
&=\begin{bmatrix}
Z_{pp}^S Y_{pp}^L + Z_{pn}^S Y_{np}^L& 0 \\ 
0 & Z_{np}^S Y_{pn}^L + Z_{nn}^S Y_{nn}^L 
\end{bmatrix} \label{eq:Lpn_semidec}
\end{align}

 The advantage of this model compared with the exact models is that the SISO-methods (e.g. Nyquist Criterion) can be applied instead of MIMO-methods (e.g. GNC) since the minor-loop gain is decoupled. The drawback of this method compared with the decoupled models (\ref{eq:Ldq_dec})-(\ref{eq:Lpn_dec}) is that the entire 2x2 impedance matrices must be identified by combining two linear independent injections.
 
\subsection{Stability analysis by eigenvalue loci}\label{sec:eigenvalues}

The main purpose of the paper is to evaluate the decoupled and semi-decoupled approximations with the exact matrix models $\mathbf{L}_{dq,\textrm{exact}}$ and $\mathbf{L}_{pn,\textrm{exact}}$. A good method for comparison is to plot the eigenvalue loci of the minor loop gains. Plotting the eigenvalue loci in the complex plane is the Nyquist plot, which is widely used for stability analysis. Obtaining the eigenvalues $\mathbf{\lambda}$ for the decoupled models is straightforward since:
\begin{equation}\label{eq:lambda_11}
eig\left\{ 
\begin{bmatrix}
L_{dd} & 0 \\0 & L_{qq}
\end{bmatrix}\right\}
=
\begin{bmatrix}
\lambda_1 \\ \lambda_2
\end{bmatrix}
=
\begin{bmatrix}
L_{dd} \\ L_{qq}
\end{bmatrix}
\end{equation}

Consequently, the eigenvalues of all decoupled models will be simply the diagonal elements in (\ref{eq:Ldq_dec})-(\ref{eq:Lpn_semidec}). On the other hand, obtaining the eigenvalue loci of the 2x2 matrix impedance models is slightly more challenging. One method is to solve numerically the following equation for each frequency:
\begin{equation}
    det\left(
    \begin{bmatrix}
    \lambda(s) & 0 \\0 & \lambda(s)
    \end{bmatrix}
    -\mathbf{L}_{dq,\textrm{exact}}(s) \right) = 0
\end{equation}

This expression can be expanded and solved for $\lambda$:
\begin{align}\label{eq:lambda_22}
    &(\lambda-L_{dd}) ( \lambda - L_{qq} ) - L_{dq}L_{qd}=0 \nonumber \\
    &\lambda^2-\lambda (L_{dd}+L_{qq}) + L_{dd}L_{qq} -  L_{dq}L_{qd}=0 \nonumber \\
    &\lambda = \frac{1}{2}\left(L_{dd}+L_{qq} \pm \sqrt{ (L_{dd}-L_{qq})^2-4L_{dq}L_{qd}} \right)
\end{align}

where the $\pm$-sign will give the two solutions for $\lambda$: $\lambda_1$ and $\lambda_2$. The corresponding equation for the sequence domain matrix $\mathbf{L}_{pn,\textrm{exact}}$ can be found by replacing $d \rightarrow p$ and $q \rightarrow n$.

\subsection{Classification of previous works}
In light of the many definitions provided earlier in this section, a classification of previous works is presented in Table \ref{tab:prev_work}. Most of the previous works in the sequence domain has applied the decoupled model (\ref{eq:Zpn_dec}) since the definition of the modified sequence domain and its 2x2 matrix is relatively new. Note that different authors use different notation for the same matrix. Most of the work in the \textit{dq}-domain has used the full matrix model (\ref{eq:Zdq_full}), but some of these papers have assumed semi-decoupled models when performing the stability analysis. Using the decoupled model in \textit{dq}-domain (\ref{eq:Zdq_dec}) has, to the authors knowledge, not been performed.
\begin{table}[ht]
    \centering
    \caption{Examples of previous works with different impedance models}
    \begin{tabular}{r|c|c}
                        & Sequence domain & \textit{dq}-domain \\ \hline
         Matrix model   & \cite{Rygg2016} \cite{Shah2016} \cite{Ren2016} \cite{Bakhshizadeh2016}  & \cite{Belkhayat1997} \cite{Francis2011} \cite{Burgos2010} \cite{Wen2015b} \cite{Liu2015} \cite{Familiant2009} \\
         Semi-decoupled &  \cite{Shah2016}   & \cite{Burgos2010} \cite{Wen2015b}   \\
         Decoupled      &  \cite{Sun2011} \cite{Cespedes2014b} \cite{Roinila2014}    &-  \\
    \end{tabular}
        \label{tab:prev_work}
\end{table}

\section{Impact of Mirror Frequency Coupling}\label{sec:MFD}
Mirror Frequency Coupling (MFC) and Mirror Frequency Decoupled (MFD) systems were defined in \cite{Rygg2016}. It will be highlighted in the present paper that MFD is a very important property when discussing the accuracy of decoupled impedance models. By definition, the modified sequence domain is decoupled for MFD systems, and the matrix $\mathbf{Z}_{pn}$ will have the following structure:
\begin{equation}\label{eq:Zpn_mfd}
   \mathbf{Z}_{pn}\Big|_{\textrm{MFD}}
   =
   \begin{bmatrix}
   Z_{p} & 0 \\ 0 & Z_{n}
   \end{bmatrix}
\end{equation}

That is, $Z_{pn}=Z_{np}=0$, $Z_{p}=Z_{pp}$ and $Z_{n}=Z_{nn}$. This leads to the following implication:

\begin{equation}\label{eq:MFDimp}
\textnormal{MFD} \implies \mathbf{Z}_{pn}=\mathbf{Z}_{pn,\textrm{dec}}=\mathbf{Z}_{pn,\textrm{semidec}}
\end{equation}

On the other hand, the \textit{dq}-domain is not decoupled, but its impedance matrix will have the following structure \cite{Rygg2016}:
\begin{equation}\label{eq:Zdq_mfd}
   \mathbf{Z}_{dq}\Big|_{\textnormal{MFD}}
   =
   \begin{bmatrix}
   Z_{\textnormal{diag}} & Z_{\textnormal{offdiag}} \\ -Z_{\textnormal{offdiag}} & Z_{\textnormal{diag}}
   \end{bmatrix}
\end{equation}

Since the \textit{dq}-domain off-diagonal elements are clearly not zero for MFD-systems, an error will be introduced when neglecting them, as is the case for both $\mathbf{Z}_{dq,\textrm{dec}}$ and $\mathbf{L}_{dq,\textrm{semidec}}$.

As discussed in \cite{Rygg2016}, there are several sources to coupling. Essentially, all sub-blocks of the system need to follow the structure of (\ref{eq:Zpn_mfd}) and (\ref{eq:Zdq_mfd}) in order to have a fully MFD system. Table \ref{tab:mfd} provides a MFD categorization of components and control system blocks.

\begin{table}[h!]
    \centering
    \caption{MFD categorisation of typical components and control system blocks}
    \begin{tabular}{c|c}
         \textbf{Mirror Frequency \underline{Decoupled}} & \textbf{Mirror Frequency \underline{Coupled} } \\ \hline
         Linear passive elements (R,L,C)   & Phase Lock Loop (PLL) \\
         \textit{dq} current controller      & DC-link voltage controller \\
         $\alpha \beta$ controllers          & Salient-pole machines \\
         Transformers and cables             & Active and reactive power controllers \\
         Round-rotor machines
    \end{tabular}
        \label{tab:mfd}
\end{table}

Linear passive elements, cables, transformers and round-rotor machines are linear in the phase domain. In other words, they will only respond at the same frequency at which they are excited. Consequently, they will not bring mirror frequency coupling to the system. This is also true for current controllers and $\alpha \beta$ controllers, since they have the symmetric structure given by (\ref{eq:Zdq_mfd}). On the other hand, the DC-link voltage controller is MFC since it only acts on the d-axis and can therefore not comply with (\ref{eq:Zdq_mfd}). Similarly, the PLL only acts on the q-axis voltage, and will therefore be MFC. Salient-pole machines are MFC since the reluctance in \textit{d}- and \textit{q}-axes differ, hence $Z_{dd}\neq Z_{qq}$. Finally, active and reactive power controllers are MFC since they are non-linear, and their linearized equivalent will depend on the operation point. They will therefore not comply with (\ref{eq:Zdq_mfd}).

\subsection{Decoupling norm $\epsilon$}

Defining a norm for measuring the degree of coupling in the impedance matrices is useful when discussing the accuracy of impedance models of a system. In \cite{Burgos2010} a norm called $AC_{index}$ was defined for this purpose. Another norm was defined in \cite{Shah2016} based on diagonal dominance. Both norms are related with the theory of Gershgorin circles, and are simple to apply, but can be too conservative. In the present paper a norm $\epsilon$ is defined based on the difference between the exact and the semi-decoupled impedance models. In the \textit{dq}-domain this is obtained by combining (\ref{eq:lambda_22}) with (\ref{eq:lambda_11}):

\begin{align}
    &\lambda_{dq,\textrm{exact}} = 
    \begin{bmatrix} L_{dd}-\epsilon_{dq} \\L_{qq}+\epsilon_{dq}\end{bmatrix}
    =\lambda_{dq,\textrm{semidec}}+\begin{bmatrix}-1\\1\end{bmatrix}\epsilon_{dq}\nonumber\\
    &\epsilon_{dq} = \frac{1}{2}\left(L_{dd}-L_{qq}-
    \sqrt{\left(L_{dd}-L_{qq}\right)^2+4L_{dq}L_{qd}}\right)\label{eq:epsilon}
\end{align}

The corresponding sequence domain norm $\epsilon_{pn}$ is obtained by replacing $d \rightarrow p$ and $q \rightarrow n$. From (\ref{eq:epsilon}) it is clear that when $\epsilon_{dq}$ or $\epsilon_{pn}$ is sufficiently small in magnitude, the corresponding semi-decoupled model will give identical eigenvalues as the exact ones. Application of these norms has been carried out by simulations in section \ref{sec:norm}.

\subsection{Impact of grid X/R-ratio on \textit{dq}-domain decoupling}\label{sec:impact_XR}
\begin{figure}[ht!]
     \centering
     \includegraphics[width=0.42\textwidth]{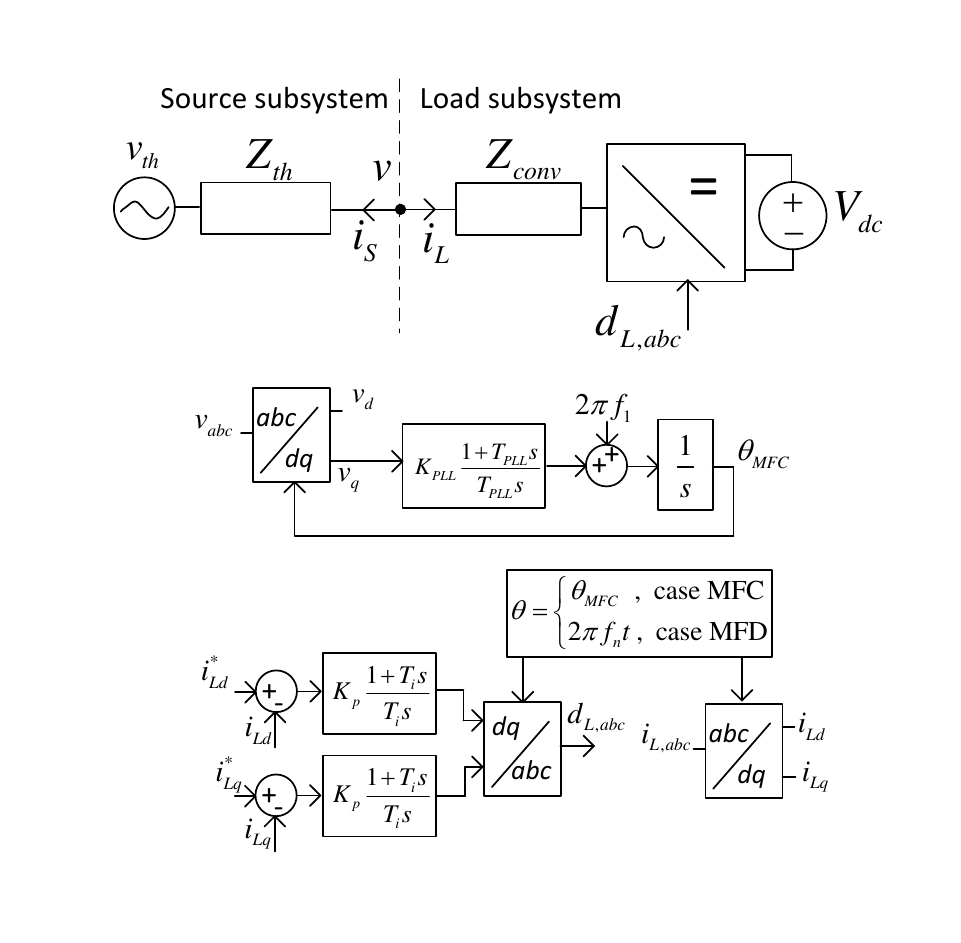}
     \caption{Detailed schematic of the simulation case system}
     \label{fig:case_schematic}
\end{figure}

One major contributor to coupling in the \textit{dq}-domain impedance matrices is the presence of inductance and capacitance. The reactance at fundamental frequency will appear in the off-diagonal elements \cite{Wen2015b}. Taking an RL-equivalent as an example (Fig. \ref{fig:case_schematic}), the \textit{dq}-domain impedance matrix will be:
\begin{equation}
\mathbf{Z}_{dq,th}=
\begin{bmatrix}
R_{th}+sL_{th} & -\omega_1 L_{th} \\ \omega_1 L_{th} & R_{th} + s L_{th}
\end{bmatrix}
\end{equation}

The condition for this matrix to be diagonally dominant is 
\begin{equation}
(\omega_1 L_{th})^2 < R_{th}^2 + (\omega L_{th})^2
\end{equation}
This is satisfied for sufficiently high frequencies $\omega$, but also for sufficiently low X/R-ratios ($X=\omega_1 L$). In one previous application of the \textit{dq}-domain semi-decoupled model \cite{Wen2015b} the X/R-ratio was 0.07, which is considered very low.

\begin{figure}[ht]
     \centering
     \includegraphics[width=0.49\textwidth]{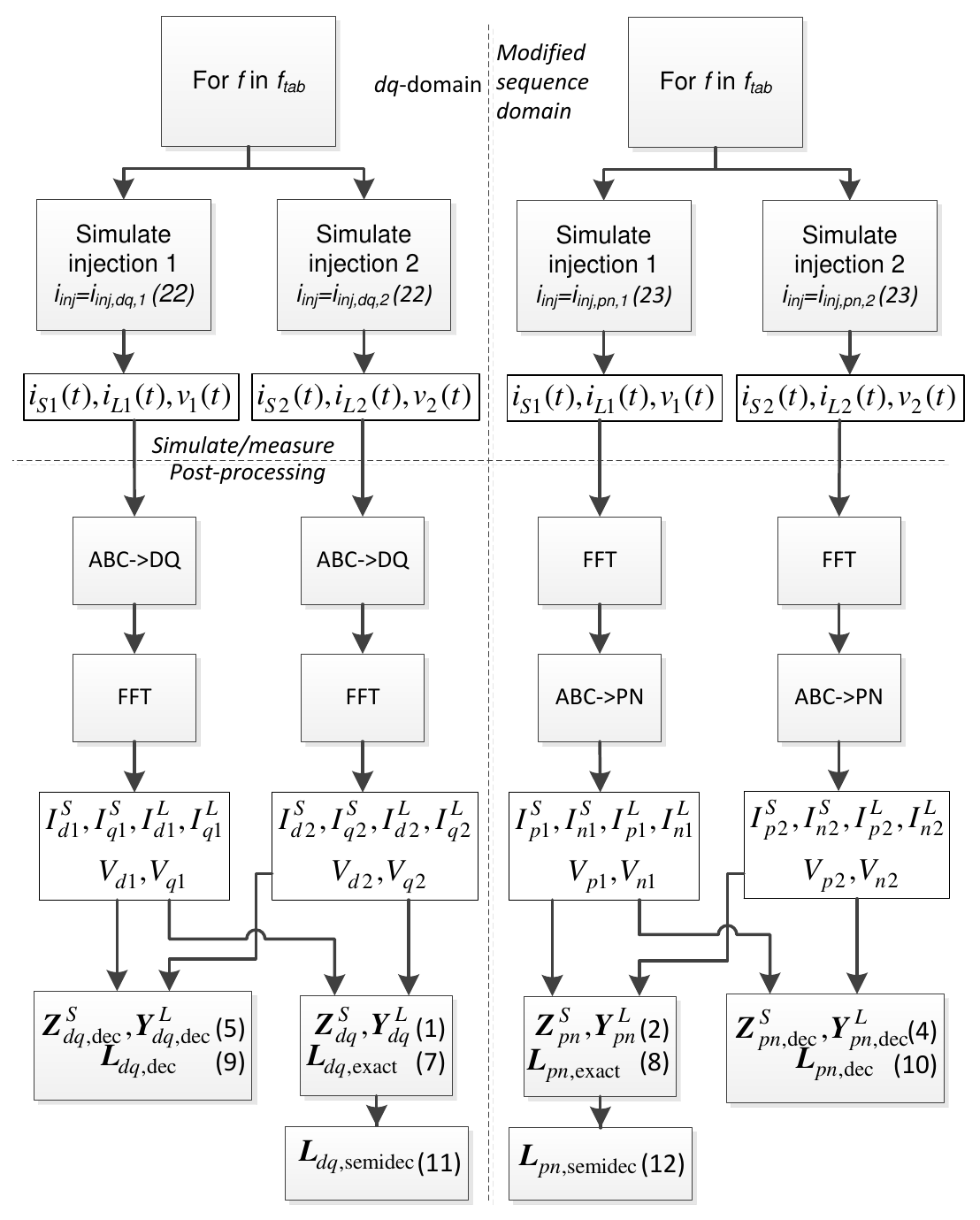}
     \caption{Flowchart for obtaining all impedance models based on simulation/measurements}
     \label{fig:flowchart}
\end{figure}

\section{Choice of injection signals}\label{sec:inj}
In this paper it is assumed that the system is time invariant in \textit{dq}-domain, and hence the 2x2 impedance matrices $\mathbf{Z}_{dq}$ and $\mathbf{Z}_{pn}$ will give exact results. However, all methods for decoupling can give errors in certain cases, and this error may or may not be sensitive to what injection signal is being applied. In previous work it was found that decoupled impedances depend on whether the injection is shunt or series \cite{Rygg2016}. In this paper, only shunt injection is assumed, but the injection signal can still be chosen in different ways. Two sets of injection signals will be considered in the analysis, $i_{inj,dq}$ and $i_{inj,pn}$. Both sets have two linear independent three-phase signals, denoted with subscript 1 and 2. This is required to establish the impedance matrices as explained in \cite{Francis2011}. 

\begin{align}
i_{inj,dq,1}(t) &= I_{inj}
\begin{bmatrix}
\sin\left(\omega_{inj}t\right)\cos\left(\omega_{1}t\right)
\\
\sin\left(\omega_{inj}t\right)\cos\left(\omega_{1}t-\frac{2\pi}{3}\right)
\\
\sin\left(\omega_{inj}t\right)\cos\left(\omega_{1}t+\frac{2\pi}{3}\right)
\end{bmatrix}
\nonumber \\
i_{inj,dq,2}(t) &= I_{inj}
\begin{bmatrix}
\sin\left(\omega_{inj}t\right)\cos\left(\omega_{1}t\right)
\\
\sin\left(\omega_{inj}t\right)\cos\left(\omega_{1}t+\frac{2\pi}{3}\right)
\\
\sin\left(\omega_{inj}t\right)\cos\left(\omega_{1}t-\frac{2\pi}{3}\right)
\end{bmatrix} \label{eq:inj_dq}
\\
i_{inj,pn,1}(t) &= I_{inj}
\begin{bmatrix}
\sin\left( \left[\omega_{inj}+\omega_1\right]t\right)
\\
\sin\left( \left[\omega_{inj}+\omega_1\right]t - \frac{2\pi}{3} \right)
\\
\sin\left( \left[\omega_{inj}+\omega_1\right]t + \frac{2\pi}{3} \right)
\end{bmatrix}
\nonumber \\
i_{inj,pn,2}(t) &= I_{inj}
\begin{bmatrix}
\sin\left( \left[\omega_{inj}-\omega_1\right]t\right)
\\
\sin\left( \left[\omega_{inj}-\omega_1\right]t + \frac{2\pi}{3} \right)
\\
\sin\left( \left[\omega_{inj}-\omega_1\right]t - \frac{2\pi}{3} \right)
\end{bmatrix}\label{eq:inj_pn}
\end{align}

\begin{figure}[b!]
     \centering
     \includegraphics[width=0.47\textwidth]{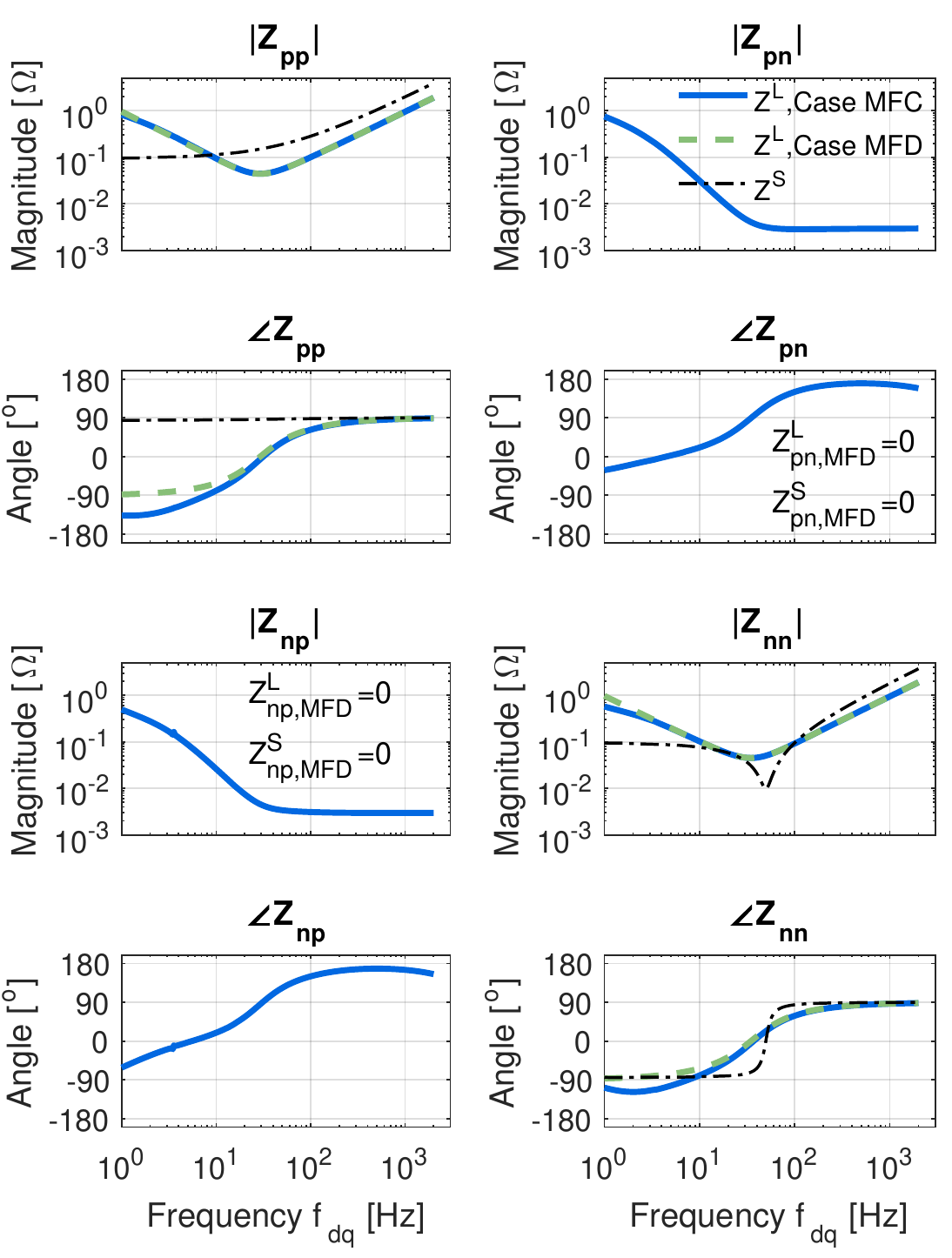}
     \caption{Magnitude/angle of $\mathbf{Z}_{pn}^S$ and $\mathbf{Z}_{pn}^L$ for both cases}
     \label{fig:Zpn_full}
\end{figure}

\begin{figure}[b!]
     \centering
     \includegraphics[width=0.49\textwidth]{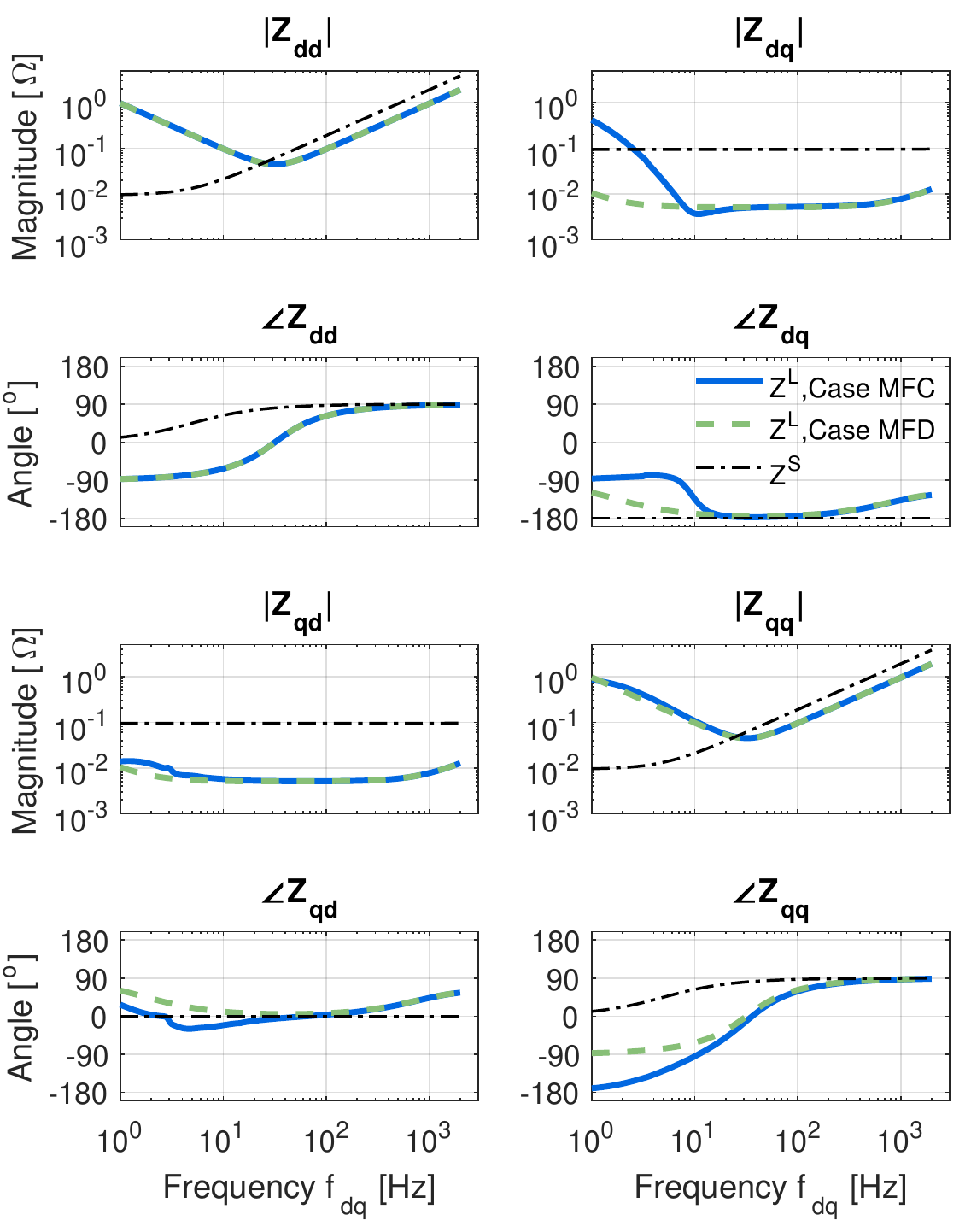}
     \caption{Magnitude/angle of $\mathbf{Z}_{dq}^S$ and $\mathbf{Z}_{dq}^L$ for both cases}
     \label{fig:Zdq_full}
\end{figure}

where $I_{inj}$ is the injection amplitude. The injection signals in the \textit{dq}-domain are defined such that $i_{inj,dq,1}$ is a pure \textit{d}-axis component, while $i_{inj,dq,2}$ is a pure \textit{q}-axis component. In the sequence domain, injection signals are defined such that $i_{inj,pn,1}$ is pure positive sequence, while $i_{inj,pn,2}$ is pure negative sequence.

Impedance matrices defined in the \textit{dq}-domain will use the \textit{dq} injection signals, while impedance matrices defined in the sequence domain will use the \textit{pn} signals.

The entire methodology has been summarized by a flowchart in Fig. \ref{fig:flowchart}. The figure is divided into two halves, one for obtaining \textit{dq}-domain impedances, and one for sequence domain impedances. Note that most of the steps are identical, the main differences are:
\begin{itemize}
    \item The injection method depends on impedance domain as explained above.
    \item The \textit{dq}-domain requires the $abc\rightarrow dq$ transform, while the sequence domain requires the symmetric component transform $abc\rightarrow pn$.
\end{itemize}

A key point to highlight is that the decoupled minor loop gains $L_{dq,dec},L_{pn,dec}$ do not require the full 2x2 impedance matrix to be established. This makes signal processing easier since the challenging off-diagonal elements shall not be identified.

\section{Simulation results}\label{sec:sim}
\subsection{Case study description}
A case study system has been defined in Fig. \ref{fig:case_schematic}. The system has been made as simple as possible in order to have illustrative results, and is a single converter connected to a Thevenin grid equivalent. The converter has a constant DC-link voltage, and controls the current in the \textit{dq}-domain according to constant set-points $i_{Ld}^*$ and $i_{Ld}^*$. The simulations are carried out by an average converter model, where the PWM is modelled by a first-order delay.

Two subcases are defined:
\begin{itemize}
\item \textbf{Case MFD}: The \textit{dq}-transform obtains its reference angle from a fixed ramp, $\theta=\omega_1t$, where $\omega_1$ is the fundamental frequency
\item \textbf{Case MFC}: The \textit{dq}-transform obtains its reference angle from a synchronous reference frame PLL, see Fig. \ref{fig:case_schematic}.
\end{itemize}

Recall from section \ref{sec:MFD} that the \textit{dq}-domain current controller is MFD, while the PLL is not. Consequently, Case MFD is truly Mirror Frequency Decoupled since there are no elements that give mirror frequency coupling. By contrast, Case MFC is not MFD due to the PLL. This will be highlighted throughout the following simulation results.

\begin{figure}[b!]
     \centering
     \includegraphics[width=0.43\textwidth]{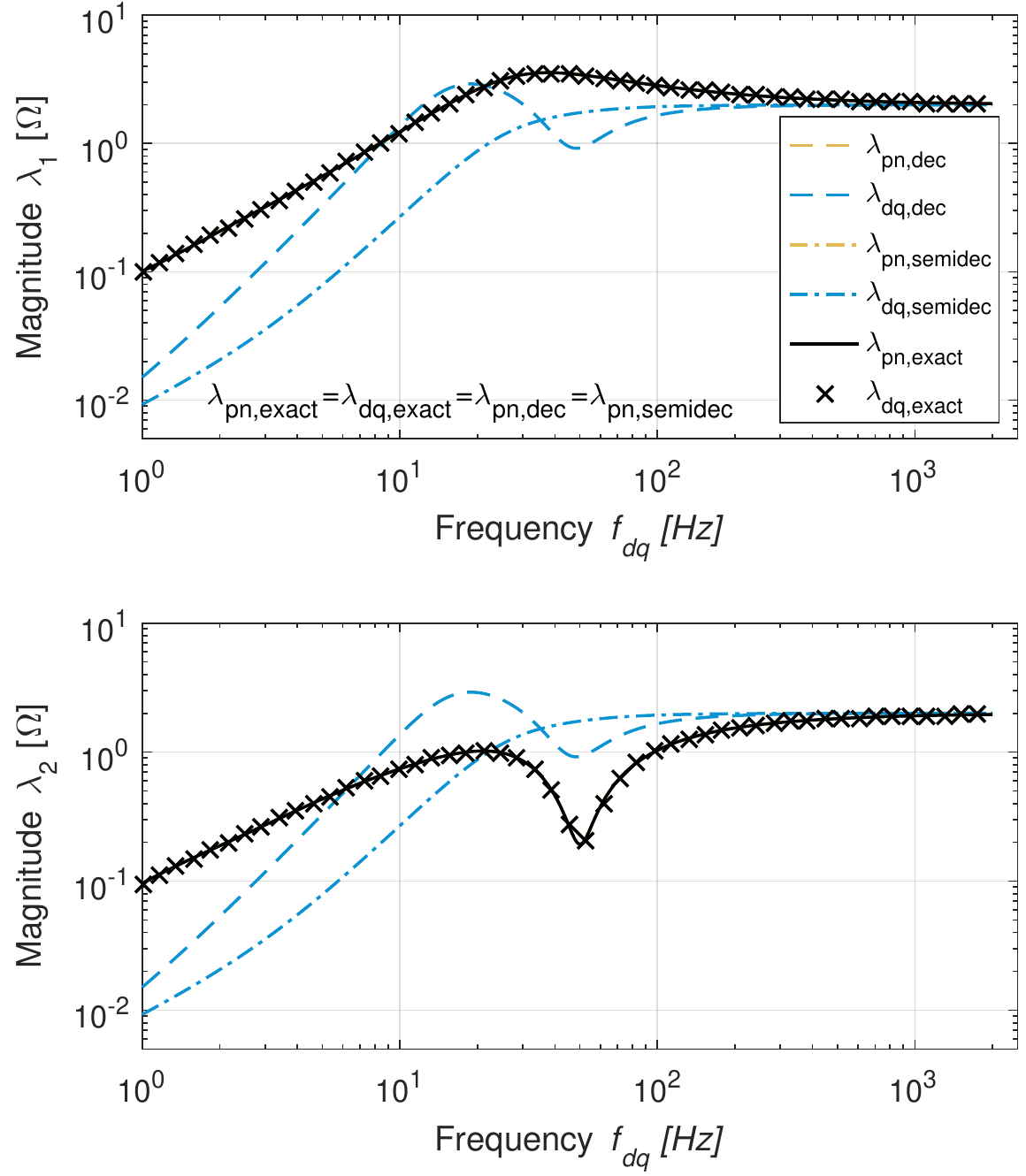}
     \caption{Comparison of eigenvalue magnitudes for case MFD}
     \label{fig:Lambda_MFD}
\end{figure}

\begin{figure}[b!]
     \centering
     \includegraphics[width=0.43\textwidth]{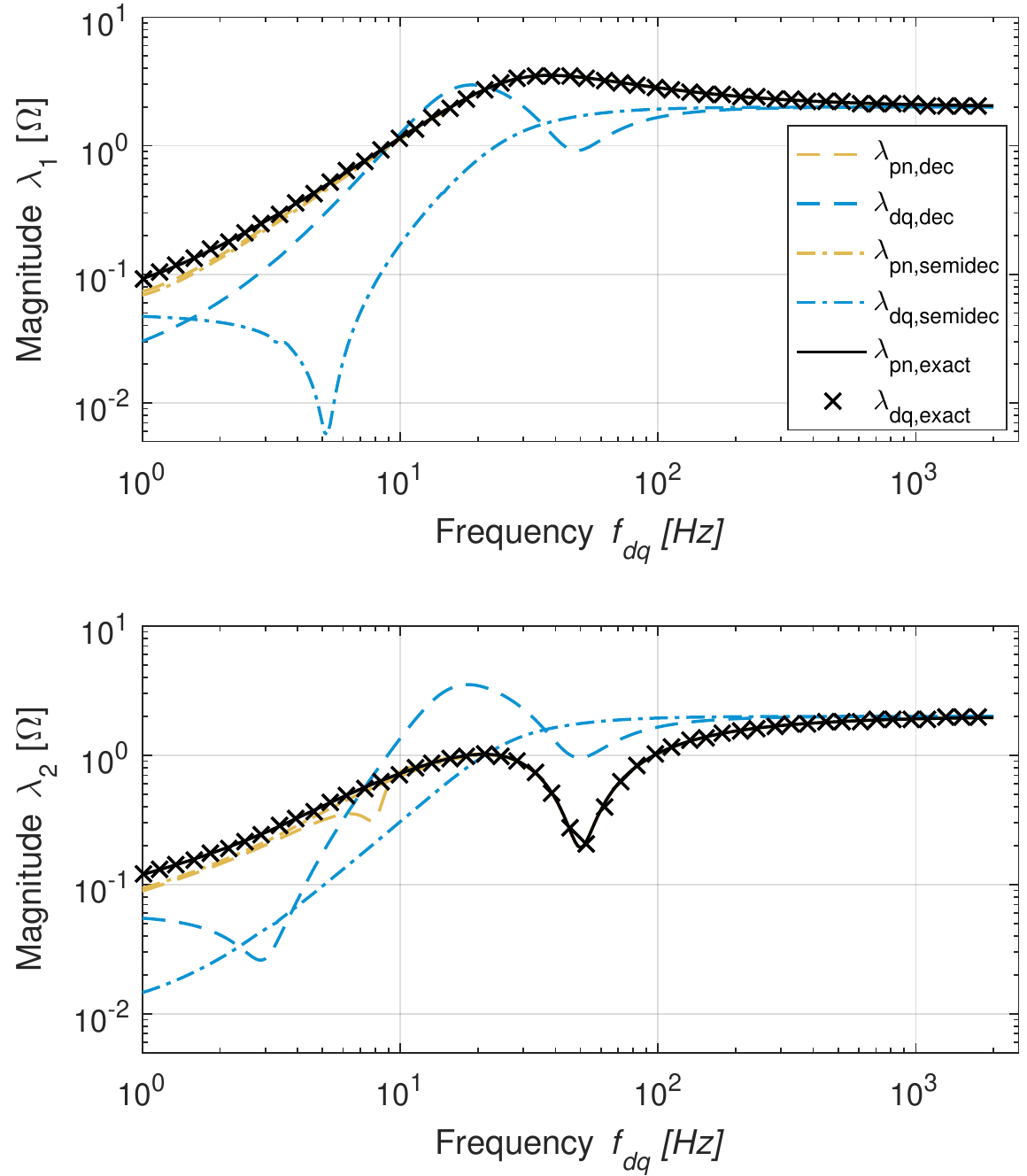}
     \caption{Comparison of eigenvalue magnitudes for case MFC}
     \label{fig:Lambda_MFC}
\end{figure}

\subsection{Evaluation of full impedance matrices $\mathbf{Z}_{dq}$ and $\mathbf{Z}_{pn}$}

The following results will explain how mirror frequency coupling impacts the impedance matrices. The modified sequence domain impedance matrix $\mathbf{Z}_{pn}$ for the two case-studies is plotted in Fig. \ref{fig:Zpn_full}. The first observation is that both the source and load impedance matrices are MFD, i.e. $Z_{pn}^S=Z_{np}^S=Z_{pn}^L=Z_{np}^L=0$. This is expected since the entire system is designed to be MFD in this case. However, in case MFC the off-diagonal elements $Z_{pn}^L$ and $Z_{np}^L$ are non-zero. This coupling is caused by the PLL, and makes interpretation and analysis more difficult.

The \textit{dq}-domain impedance matrix $\mathbf{Z}_{dq}$ is plotted in Fig. \ref{fig:Zdq_full}. There are several important differences compared with Fig. \ref{fig:Zpn_full}. First, the source subsystem impedance matrix $\mathbf{Z}_{dq}^S$ is not diagonal, and has coupling elements with magnitude $|Z_{dq}^S|=|Z_{qd}^S|=\omega_1 L_{th}$. Also, the load impedance off-diagonal elements $Z_{pn}^L$ and $Z_{np}^L$ are nonzero in both cases. It is seen that $Z_{dq}^L=-Z_{qd}^L$ and $Z_{dd}^L=Z_{qq}^L$ for Case MFD, both of which are consistent with (\ref{eq:Zdq_mfd}). In Case MFC, the \textit{dq}-domain matrix has lost its symmetric properties. $Z_{dd}$ is unchanged, while $Z_{qq}$ has a reduced angle at low frequencies. The off-diagonal element $Z_{dq}^L$ has increased significantly in magnitude, while $Z_{qd}$ is close to unchanged.

It is known that angles outside the range $[-90,90]$ are associated with positive damping, and will deteriorate the system stability. However, this principle must be applied with care for matrix impedance models, since the impact of off-diagonal elements will complicate the analysis. They can contribute to both positive and negative damping, depending on their angles with respect to the diagonal elements.

\begin{figure}[t!]
     \centering
     \includegraphics[width=0.38\textwidth]{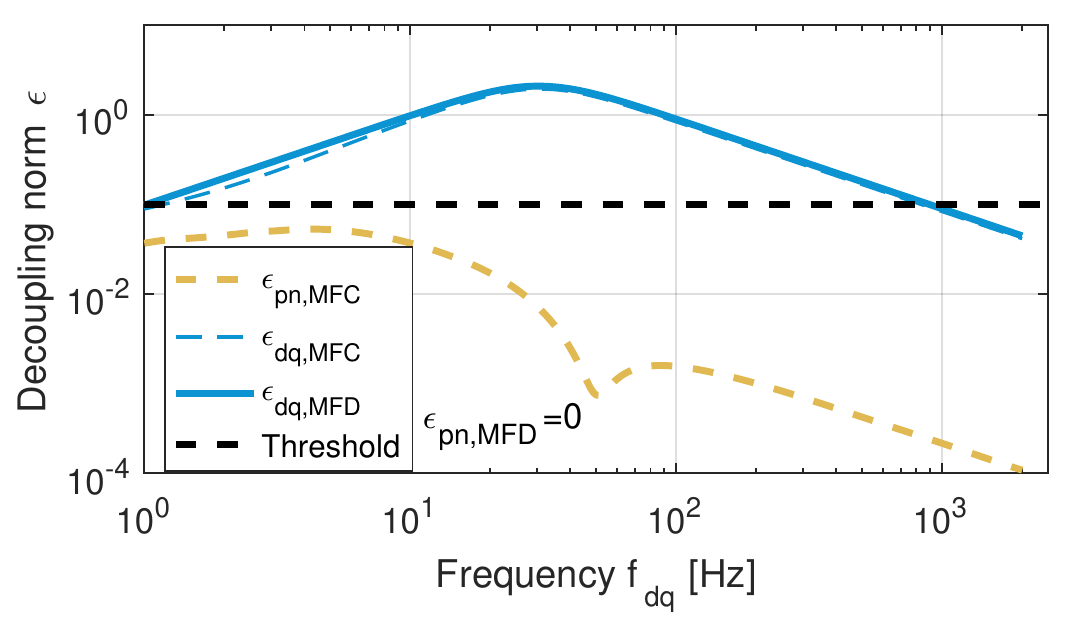}
     \caption{Overview of decoupling norms $\epsilon_{dq}$ and $\epsilon_{pn}$ for the two cases. Example threshold of 0.1 is used.}
     \label{fig:norm}
\end{figure}

\begin{figure}[t!]
     \centering
     \includegraphics[width=0.38\textwidth]{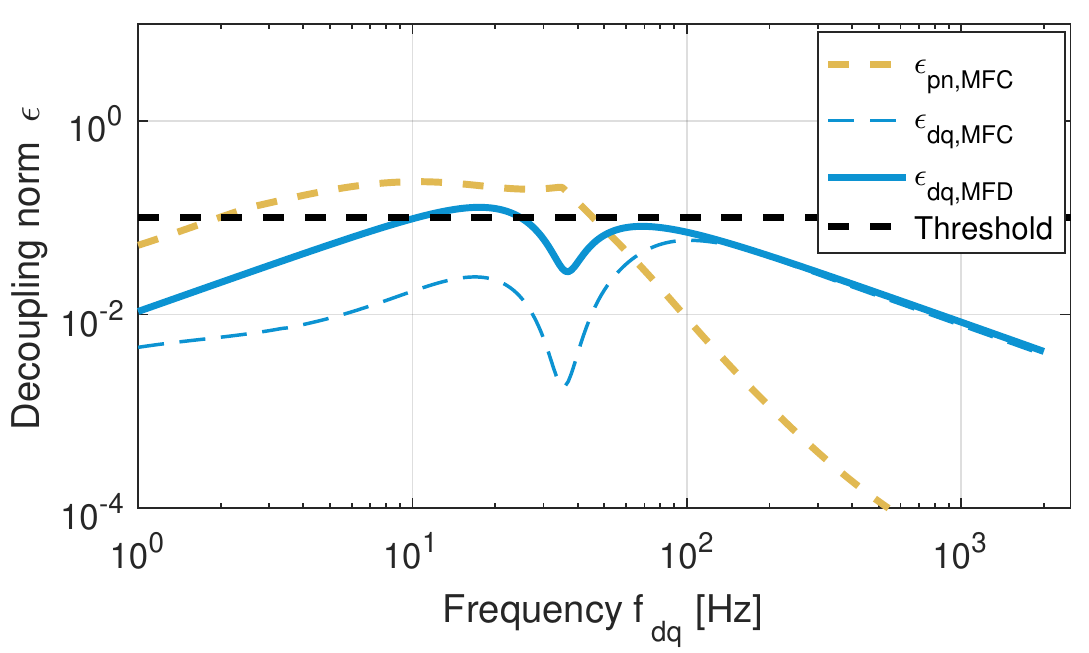}
     \caption{Decoupling norms for special case with highly resistive grid (X/R=0.1)}
     \label{fig:norm_resistive}
\end{figure}

\subsection{Comparison of eigenvalues}\label{sec:lambda_sim}

The following simulation results show a comparison of eigenvalue loci obtained from all impedance models presented in this paper. The eigenvalue loci is obtained as explained in section \ref{sec:eigenvalues}.

Fig. \ref{fig:Lambda_MFD} shows the eigenvalue loci for Case MFD. The most important observation in this figure is that all methods based on the modified sequence domain give exact results. This is consistent with (\ref{eq:MFDimp}), and underlines that matrix impedance models are not required for MFD systems in the sequence domain. On the other hand, the \textit{dq} decoupled impedance and the \textit{dq} semi-decoupled impedance do not give exact results and the error is considered large.

Fig. \ref{fig:Lambda_MFC} shows the eigenvalue loci for Case MFC. In this case none of the decoupled or semi-decoupled models give exact results for all frequencies. Again, the decoupled and semi-decoupled method in the \textit{dq}-domain give a poor approximation at frequencies up to 300 Hz.

\subsection{Comparison of decoupling norms $\epsilon_{dq}$ and $\epsilon_{pn}$}\label{sec:norm}

The decoupling norms defined in section \ref{sec:MFD} have been calculated based on the simulation results, and are plotted in Figure \ref{fig:norm}. The norm is defined as the difference between the exact eigenvalues $\lambda_{\textrm{exact}}$ and the semi-decoupled ones $\lambda_{\textrm{semidec}}$. An example threshold of 0.1 is used as an indicator for when the decoupling can be assumed without significant loss of accuracy. Note that $\epsilon_{pn,\textrm{MFD}}$ is zero by definition. 

In Case MFC, the decoupling norm in the \textit{pn}-domain lies always below the threshold of 0.1. This is consistent with Fig. \ref{fig:Lambda_MFC} where the difference between $\lambda_{pn,\textrm{semidec}}$ ando $\lambda_{pn,\textrm{exact}}$ is always below 0.1. On the other hand $\epsilon_{dq}$ violates the example threshold for all frequencies up to 1 kHz, both in the MFC and the MFD case. This is consistent with Fig. \ref{fig:Lambda_MFD} and Fig. \ref{fig:Lambda_MFC} where it is clear that $\lambda_{dq,\textrm{semidec}}$ is a poor approximation.

\subsection{Special case with highly resistive grid (X/R=0.1)}
Based on the discussion in section \ref{sec:impact_XR}, the system is analyzed for a special case with a very low X/R-ratio. Parameter values are identical to the previous case apart from $R_{th}$ and $L_{th}$. The magnitude of $Z_{th}$ is unchanged, but the X/R-ratio is changed from 10 to 0.1. The decoupling norms are plotted in Fig. \ref{fig:norm_resistive}, and it is clear that the \textit{dq}-domain is more decoupled than the sequence domain for both cases up to $\approx 70$ Hz. This is due to the fact that both subsystems in the \textit{dq}-domain are now close to decoupled. On the other hand, the sequence domain is actually less decoupled in this case compared with Fig. \ref{fig:norm}. This is due to the changes in impedance angle in the source subsystem matrix.

\section{Discussions and recommendations}\label{sec:discussion}

\subsection{Choice of impedance domain}
The modified sequence domain has normally smaller magnitude in the off-diagonal elements than the \textit{dq}-domain, and the decoupling assumption is normally best justified there.

The modified sequence domain does not require a reference transformation angle, only the fundamental frequency needs to be known in order to capture the mirror frequency components. Since the frequency is assumed constant throughout the system, it is easier to combine impedance models in the modified sequence domain by series/parallel connection. This is difficult in the \textit{dq}-domain since the reference angle depends on location.

The main advantage of the \textit{dq}-domain is that many control system blocks as well as electrical machine models, are realized in \textit{dq}-coordinates. It can therefore be easier to derive and linearize analytical impedance models in this domain. For impedance analysis based on analytic expressions, it will be a good alternative to derive models in \textit{dq}-domain, and then transform them into the modified sequence domain by (\ref{eq:imp_transf}) before doing the stability analysis.

\subsection{Decoupled vs. semi-decoupled models}
The decoupled and semi-decoupled models in the \textit{dq}-domain are generally not recommended. They are neglecting the coupling between \textit{d}- and \textit{q}-axis caused by e.g. inductance and capacitance. These models will only be valid at high frequencies, e.g. several times the fundamental. Alternatively, in cases with very low grid inductance, as presented in Fig. \ref{fig:norm_resistive}.

When choosing between decoupled or semi-decoupled models in the sequence domain, it is important to highlight the added complexity in obtaining semi-decoupled models by simulation or measurements. They require the entire 2x2-matrix to be established, and this is significantly more challenging than obtaining the decoupled equivalents by scalar equations (\ref{eq:Zpn_dec}). By the case examples in this paper the semi-decoupled models do not seem to improve the accuracy of the eigenvalue loci significantly, hence there is seemingly little advantage in applying them. The recommendation is therefore to apply decoupled models rather than semi-decoupled ones.

The final and most difficult aspect is to provide clear recommendations for when matrix (exact) models are needed. The norm $\epsilon$ has been defined for this purpose. For a simpler and conservative analysis, it is sufficient to identify the contributors to Mirror Frequency Coupling in the system. Furthermore, to estimate the frequency range where these contributors introduce coupling. Outside this frequency range, the decoupled sequence domain model will be accurate.

\section{Conclusions}\label{sec:conclusion}
This paper has provided a systematic overview and analysis of available impedance models, including a thorough discussion on their accuracy and applicability. The analysis is conducted under the assumption of time invariance in the \textit{dq}-domain. Both the \textit{dq}-domain and the sequence domain are considered. The following three terms are defined in both impedance domains:
\begin{itemize}
\item \textit{Decoupled models} initially assumes two independent SISO-models, i.e. neglects all coupling
\item \textit{Semi-decoupled models} captures all coupling by a MIMO-model, but neglects the resulting coupling at the final stage (stability analysis)
\item \textit{Exact models} represents the system by 2x2 matrices, and performs stability analysis by MIMO-methods
\end{itemize}

In addition to the above definitions, the paper contributes with the following key results:
\begin{itemize}
\item The decoupled and semi-decoupled models in the \textit{dq}-domain will have poor accuracy for many systems. They do not capture the coupling between d- and q-axis in e.g. inductance and capacitance. 
\item In the special case with a highly resistive grid (close to zero inductance), the semi-decoupled model in the \textit{dq}-domain will be accurate.
\item The decoupled and semi-decoupled models in the sequence domain give exact results by definition if both subsystems are MFD (\ref{eq:MFDimp})
\item The norm $\epsilon$ is defined to measure the error in semi-decoupled models compared with the exact models (\ref{eq:epsilon})
\item Based on the simulation case-studies, the decoupled models seem to have similar accuracy as the semi-decoupled ones. They are therefore preferred since they are significantly easier to obtain.
\end{itemize}

\small
\bibliographystyle{IEEEtran}
\bibliography{PhD-arbeid}

\begin{thebibliography}{10}
\providecommand{\url}[1]{#1}
\csname url@samestyle\endcsname
\providecommand{\newblock}{\relax}
\providecommand{\bibinfo}[2]{#2}
\providecommand{\BIBentrySTDinterwordspacing}{\spaceskip=0pt\relax}
\providecommand{\BIBentryALTinterwordstretchfactor}{4}
\providecommand{\BIBentryALTinterwordspacing}{\spaceskip=\fontdimen2\font plus
\BIBentryALTinterwordstretchfactor\fontdimen3\font minus
  \fontdimen4\font\relax}
\providecommand{\BIBforeignlanguage}[2]{{%
\expandafter\ifx\csname l@#1\endcsname\relax
\typeout{** WARNING: IEEEtran.bst: No hyphenation pattern has been}%
\typeout{** loaded for the language `#1'. Using the pattern for}%
\typeout{** the default language instead.}%
\else
\language=\csname l@#1\endcsname
\fi
#2}}
\providecommand{\BIBdecl}{\relax}
\BIBdecl

\bibitem{Belkhayat1997}
M.~Belkhayat, \emph{Stability criteria for AC power systems with regulated
  loads}.\hskip 1em plus 0.5em minus 0.4em\relax Purdue University, 1997.

\bibitem{Sun2011}
J.~Sun, ``Impedance-based stability criterion for grid-connected inverters,''
  \emph{Power Electronics, IEEE Transactions on}, vol.~26, no.~11, pp.
  3075--3078, Nov 2011.

\bibitem{Rygg2016}
A.~Rygg, M.~Molinas, Z.~Chen, and X.~Cai, ``A modified sequence domain
  impedance definition and its equivalence to the dq-domain impedance
  definition for the stability analysis of ac power electronic systems,''
  \emph{IEEE Journal of Emerging and Selected Topics in Power Electronics},
  vol.~PP, no.~99, pp. 1--1, 2016.

\bibitem{Francis2011}
G.~Francis, R.~Burgos, D.~Boroyevich, F.~Wang, and K.~Karimi, ``An algorithm
  and implementation system for measuring impedance in the d-q domain,'' in
  \emph{Energy Conversion Congress and Exposition (ECCE), 2011 IEEE}, Sept
  2011, pp. 3221--3228.

\bibitem{Desoer1980}
C.~Desoer and Y.-T. Wang, ``On the generalized nyquist stability criterion,''
  \emph{Automatic Control, IEEE Transactions on}, vol.~25, no.~2, pp. 187--196,
  Apr 1980.

\bibitem{Wen2016b}
B.~Wen, D.~Boroyevich, R.~Burgos, P.~Mattavelli, and Z.~Shen, ``Inverse nyquist
  stability criterion for grid-tied inverters,'' \emph{IEEE Transactions on
  Power Electronics}, vol.~PP, no.~99, pp. 1--1, 2016.

\bibitem{Burgos2010}
R.~Burgos, D.~Boroyevich, F.~Wang, K.~Karimi, and G.~Francis, ``On the ac
  stability of high power factor three-phase rectifiers,'' in \emph{Energy
  Conversion Congress and Exposition (ECCE), 2010 IEEE}.\hskip 1em plus 0.5em
  minus 0.4em\relax IEEE, 2010, pp. 2047--2054.

\bibitem{Wen2015b}
B.~Wen, D.~Dong, D.~Boroyevich, R.~Burgos, P.~Mattavelli, and Z.~Shen,
  ``Impedance-based analysis of grid-synchronization stability for three-phase
  paralleled converters,'' \emph{Power Electronics, IEEE Transactions on},
  vol.~PP, no.~99, pp. 1--1, 2015.

\bibitem{Shah2016}
S.~Shah and L.~Parsa, ``Sequence domain transfer matrix model of three- phase
  voltage source converters,'' in \emph{IEEE PES General Meeting, Boston
  US}.\hskip 1em plus 0.5em minus 0.4em\relax IEEE, 2016.

\bibitem{Ren2016}
W.~Ren and E.~Larsen, ``A refined frequency scan approach to sub-synchronous
  control interaction (ssci) study of wind farms,'' \emph{IEEE Transactions on
  Power Systems}, vol.~31, no.~5, pp. 3904--3912, Sept 2016.

\bibitem{Bakhshizadeh2016}
M.~K. Bakhshizadeh, X.~Wang, F.~Blaabjerg, J.~Hjerrild, L.~Kocewiak, C.~L. Bak,
  and B.~Hesselbæk, ``Couplings in phase domain impedance modeling of
  grid-connected converters,'' \emph{IEEE Transactions on Power Electronics},
  vol.~31, no.~10, pp. 6792--6796, Oct 2016.

\bibitem{Liu2015}
Z.~Liu, J.~Liu, W.~Bao, and Y.~Zhao, ``Infinity-norm of impedance-based
  stability criterion for three-phase ac distributed power systems with
  constant power loads,'' \emph{Power Electronics, IEEE Transactions on},
  vol.~30, no.~6, pp. 3030--3043, June 2015.

\bibitem{Familiant2009}
Y.~Familiant, J.~Huang, K.~Corzine, and M.~Belkhayat, ``New techniques for
  measuring impedance characteristics of three-phase ac power systems,''
  \emph{Power Electronics, IEEE Transactions on}, vol.~24, no.~7, pp.
  1802--1810, July 2009.

\bibitem{Cespedes2014b}
M.~Cespedes and J.~Sun, ``Impedance modeling and analysis of grid-connected
  voltage-source converters,'' \emph{Power Electronics, IEEE Transactions on},
  vol.~29, no.~3, pp. 1254--1261, March 2014.

\bibitem{Roinila2014}
T.~Roinila, M.~Vilkko, and J.~Sun, ``Online grid impedance measurement using
  discrete-interval binary sequence injection,'' \emph{Emerging and Selected
  Topics in Power Electronics, IEEE Journal of}, vol.~2, no.~4, pp. 985--993,
  Dec 2014.

\end{thebibliography}

\appendix

\subsection{Parameter values used in simulations}
\label{app:params}
\begin{table}[h!]
    \centering
    \begin{tabular}{l | l | l}
        $V_{th}$=  690 V   &   $S_{base}$ = $1$ $MW$ & $f_{n}$=  $50$ $Hz$\\
        $V_{dc}$=  $1400$ $V$   & $Z_{th}$ = $0.02+j0.4$ $p.u.$   &   $Z_{S}$ = $0.002+j0.1$ $p.u.$ \\
        $K_{p}$ = $0.255$ $p.u.$  & $T_{i}$=  $0.0025$ $s$ &  \\
        $K_{PLL}$ = $60$ $s$  &  $T_{PLL}$ = $0.033$ $s.$ &   \\
    \end{tabular}
    \caption{Parameter values applied in the simulation cases}
    \label{tab:params}
\end{table}

\ifCLASSOPTIONcaptionsoff
  \newpage
\fi

\end{document}